# Mobile Testbeds with an Attitude

Sungwook Moon, Ahmed Helmy
Department of Computer and Information Science and Engineering
{sungwook, helmy}@ufl.edu

## ABSTRACT
There have been significant recent advances in mobile networks, specifically in multi-hop wireless networks including DTNs and sensor networks. It is critical to have a testing environment to realistically evaluate such networks and their protocols and services. Towards this goal, we propose a novel, mobile testbed of two main components. The first consists of a network of robots with personality-mimicking, human-encounter behaviors, which will be the focus of this demo. The personality is build upon behavioral profiling of mobile users based on extensive wireless-network measurements and analysis. The second component combines the testbed with the human society using a new concept that we refer to as participatory testing utilizing crowd sourcing.

## 1. INTRODUCTION
Testbeds have been an important part of infrastructure networks for testing communication protocols in real-world environments. Realistic testbed environments provide opportunities not only to fine-tune parameters but to also find unexpected errors that could not have been observed in simulation. In DTN, many of the protocols and mobility models have been evaluated via simulations; however, standard testbed environments are not well established. Mobile networks (including handheld devices) are becoming very personal and tightly coupled with users, hence capturing behavioral characteristics is crucial for the evaluation of future mobile networking protocols, architectures and services. This behavioral profile can be captured using personalities. To achieve this goal, we propose a novel testbed architecture that incorporates the networks of robots with personality and participatory testing by utilizing crowd sourcing. Specifically, networks of robots have personalities that act based on their behavioral profiles. They communicate via mobile devices with other robots and humans. Robots carry the same types of mobile devices as humans carry, and these mobile devices act as personalities for the robots. Therefore, portability is one of the strengths of this architecture for networks of robots. Other advantages are a realistic testing environment and a scalable testbed through participatory testing, achieved by using human society as a testing environment. Our approach is novel in that we 1) implant personality in robots to mimic human mobility; and 2) promote scalable communication by recruiting humans as test subjects to create a realistic and flexible testbed environment.

## 2. TESTBED DESIGN
Robots, like humans, have personalities and behave accordingly. This makes the robot equal to the human in terms of mobility behavior. Both the robots and humans carry mobile devices, but the devices that the robots carry have one more interfaces that interact with the robots that carry them. This personality interface gives movement instructions and movement orders and receives sensor readings to control the robot. Figure 1 illustrates this communication structure. Communication protocol is implemented and performed by mobile devices separately from the personality interface. Personality can also trigger mobility through interaction with communication protocol.

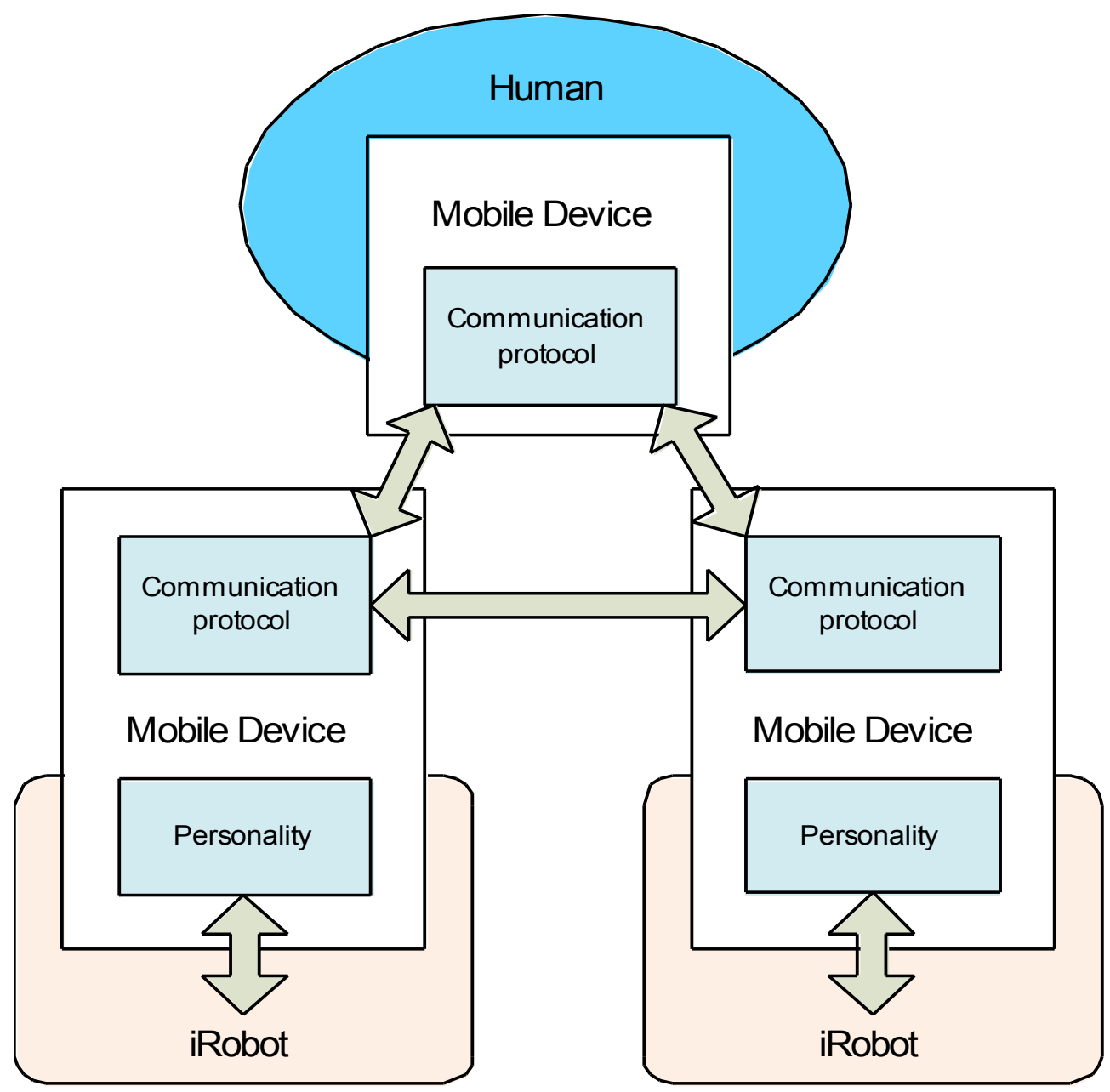

**Figure 1. Communication structure among robots, personality interface, and communication protocol**

### 2.1 Robot with personality.

#### *2.1.1 Mobility generation from encounter patterns*
The robot moves according to its behavioral profile emulating a human being. This requires a deep understanding of human mobility. As the robots are not aware of location information, they need to use encounter information for mobility. This also fits the purpose of robot use, as the objective of the communication protocol is message exchange with encountered target nodes instead of locations. To generate mobility from a human encounter pattern, Whitbeck et al. [7] proposed methodologies to infer mobility trace from the encounter trace. We adapted a similar approach for basic robots movements that resemble human encounters. Robots have the following components for movement: attraction, repulsion, and drag. Attraction is an intention to move towards other nodes; repulsion is a force to move away from particular nodes; drag is a property to stay and not to be affected by other nodes. Each character has different forces for different nodes. These parameters are based on the encounter trace from real-world measurements such as Bluetooth encounter or WLAN trace.

### *2.1.2 Applying periodical encounter behavior*

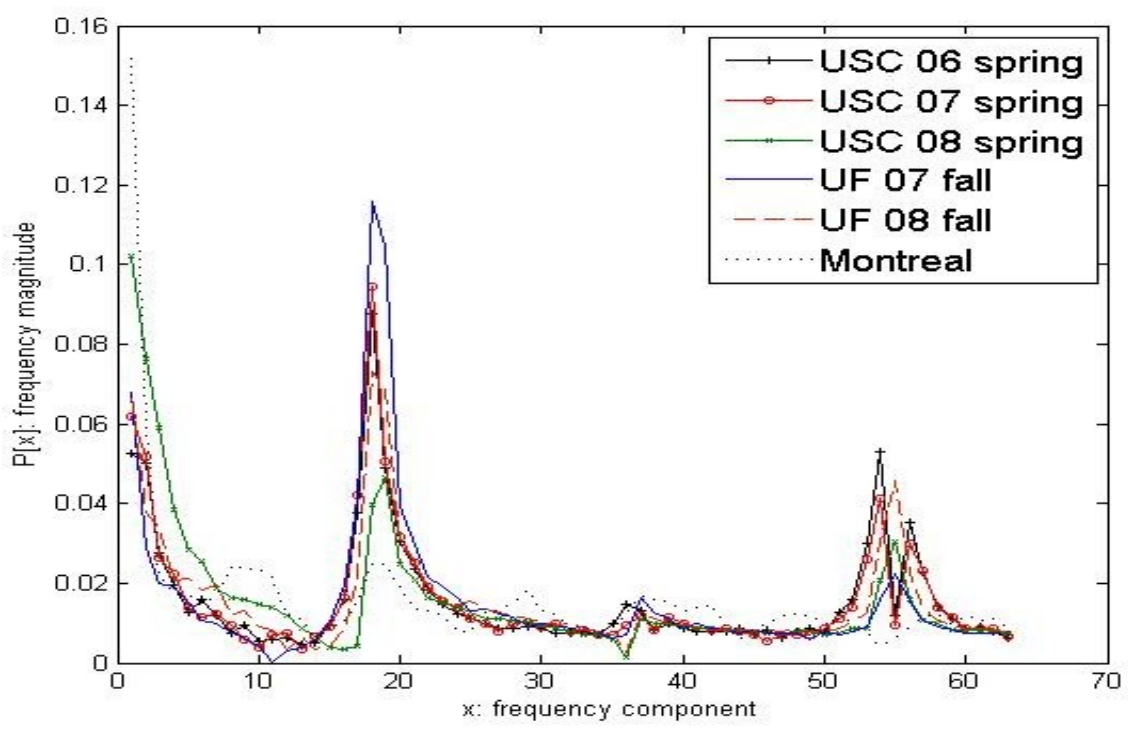

**Figure 2. Spectral analysis of mobile users to identify the periodicities in their encounter pattern**

In our recent study [5] of human encounters shown in Figure 2, we applied spectral analysis on the WLAN trace for 128 days to observe periodicity in mobile users' encounters. In the figure, each frequency component indicates a number of repeated patterns and frequency magnitude indicates how significant the pattern is. We can see the peak at a frequency component of around 18. This frequency component corresponds to a 7-day cycle as it means a specific pattern has repeated 18 times over 128 days, which translates to every 7 days (128/18=7.x). For most of the encountered pairs, we observed strong periodicity, and particularly a weekly encounter. We reflect these periodic properties on the robots to emulate human encounter behavior. Specifically, robots attempt to encounter certain nodes periodically depending on the purpose of experiments. Periodicity can vary by its frequency component (e.g. an encounter every 10 mins). Periodicity can be composed of ,multiple frequency components. (e.g., an encounter every 10 mins. and every 35 mins) Although these periodic properties enable robots to mimic human encounter patterns, further investigation is needed to validate the accuracy of this approach in reproducing the main statistical characteristics of the traces.

## 2.2 Participatory testing

Encounter measurements have been performed by multiple research projects such as MIT's Reality Mining [9] and Haggle projects [10]. Yet, the total numbers of participants was limited to less than a hundred. Researchers used WLAN trace as an indirect way of measuring encounters [6] because of its large sample size. We overcome this scalability issue by publicly recruiting human subjects. Specifically, we create a public community for the testbed so users can download the programs to participate in the test. These participants form a large testbed. To achieve this goal, it is essential to develop a programmable interface to make the testing easier. We implemented communication programs on the Windows Mobile-based (HP iPAQ) and Linux-based MAEMO (Nokia N810) smart phones.

Many scenarios and protocols can be tested through this participatory testing, including the DTN routing protocols based on social models. The Profile-cast [1] is one of such protocols that provide communication based on the behavioral profiles of humans via social sensing. A message is routed based on user profiles as a function of the users' mobility preferences. The Profile-cast was tested with students in computer networking class in 2010 spring. However, the system has yet to be tested in a public forum, which is not our focus of this demonstration. We illustrate the integration of robots and human networks in figure 3.

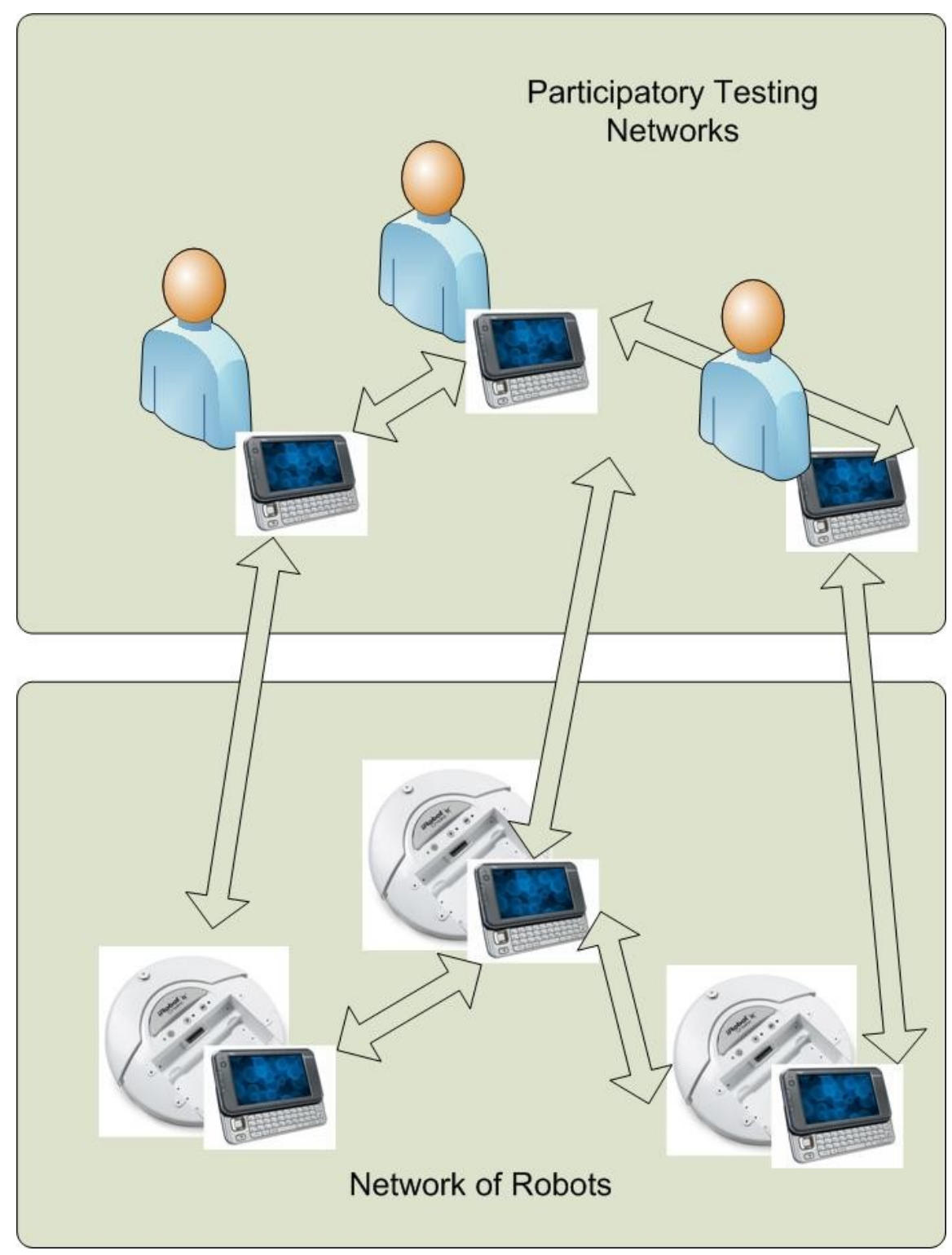

**Figure 3. Framework of testbeds for communication between networks of robots and participatory testing networks**

# 3. METHODOLOGY

## 3.1 Controlling iRobot

For easy programming and wide deployments of the device, we choose the iRobot to form the networks of robots. The connection between a mobile device (e.g,. HP iPAQ) and the iRobot is performed via Bluetooth communication. In this way, the iRobot can be controlled remotely. The personality interface can send commands and receive sensor readings to/from iRobot. For compatibility and portability, we use mobile devices instead of laptops. As iRobot does not control itself but is rather controlled by the personality in a separate computer (mobile device), it is the personality interface that decides the next behavior of the iRobot upon receiving signal readings from the iRobot or a communication on the mobile devices.

### 3.2 Wireless communication

The proposed structure can incorporate any communication methods/protocols by using the mounted iRobot controller as a communication device. The controller can communicate via Bluetooth or WLAN for message delivery and profile exchange. In the Profile-cast implementation, we use WLAN for collecting location information to create profiles and Bluetooth for message exchanges.

### 3.3 Personality

We create a personality on the iRobot by setting up a few behavioral rules. Various personalities can be created depending on the scenarios, protocols, and experimental environments. Their movements are achieved by the three rules mentioned earlier: attraction, repulsion and draw. Upon encountering the desired target, iRobot recognizes it as a virtual wall and stops. If the target node moves away, the personality interface triggers the attraction so that it can get closer to the target node. Repulsion is triggered to get away from the currently encountering node when the iRobot has spent enough time according to the given personality profile.

## 4. PRESENTATION SCENARIO

We create an environment where the iRobot move around with given set of instructions to reflect the behavioral profile. For communication protocol, we implement Profile-cast that show two types of delivery modes: 1) to deliver message bundles to target nodes with matching behavioral profiles, or 2) to disseminate messages to the interested nodes when such profiles are unavailable. Demonstrating a protocol such a Profile-cast on a large scale is our goal for the testbed; however, it requires a behavioral profile collection step, which associates the nodes with location-visiting preferences. We therefore modify the Profile-cast protocol to fit in the conference environment, yet still reveal the full concept of Profile-cast. Our demonstration scenario is as follows:

1) Selected users carry the mobile devices. These devices collect the encounter trace for the mobile carrier.
2) Based on the collected traces, we plant a personality profile to the iRobots mimicking each user.
3) Mobile devices contain Profile-cast implementation, and message exchange is tested.
4) Evaluate the results after the experiment is over.

## 5. RELATED WORK

MiNT [2] is a miniaturized network testbed that solely uses iRobots in a controlled space. A server computer controls the movements and communication amongst iRobots that are equipped with WLAN. Although this testbed can expand with multiple numbers of iRobots and be effective in experiments for small-scale mobile adhoc networks, it still suffers from scalability and a diversity of nodes. Roomba MADNeT [3] showed the capability of using iRobot for DTN. The researchers mounted a wireless router that runs on Linux by connecting through modified serial cable. This process might take advantage of a costumed lightweight programming board to utilize the wireless communication feature specifically for the testing purpose. However, this process can be tedious and cumbersome to many researchers who are not skilled in this area. Our method is simple and uses the existing device. Connection requires a minimum of steps and effort: either Bluetooth pairing or connecting a serial cable directly with a distant or attached computer. In [4], the authors proposed a DTN testbed. They used enclosures to contain the laptop computers and measure the signal attenuation for implemented DTN protocol. The design is centralized and users can view the wireless nodes moving around from the server computer. Mobility is limited to a controlled environment, as the participants are to follow the given paths and required to be in the experiment range. In our presentation, nodes can have complete control of their movements, including message propagation decisions. Moreover, measurements are decentralized in our experiment, as each measurement record is kept inside the mobile nodes. SCORPION [8] is a heterogeneous networking testbed, which uses iRobots, Buses, Aircrafts and humans with briefcase nodes. It provides a testbed to experiment communication between diverse movements; however, mobility of all the mobile nodes are limited to controlled movements; thus, social aspects are not reflected. Our unique contribution comes from the autonomous robots with behavioral profiles and participatory humans that provides uncontrolled, thus, realistic testing environment.

## 6. FUTURE WORK

Personality may change the efficiency of communication significantly. In our earlier investigation [5], periodical encounter appeared to be a prevalent trend among mobile nodes. We are in the process of reflecting the results of analysis on the robot personalities. In addition, we plan to deploy a large-scale experiment for DTN communication with the help of participatory DTN. Small-scale experiments should precede a large-scale experiment to minimize the time-consuming process for identifying the implementation problems. We will address the issues of recruiting the human participants and utilizing robots in a larger scale experiment by moving forward step-by-step. Under this study, we have plans to deploy a disaster recovery network of mobile nodes that cooperate with each other via Profile-cast.